\documentclass[12pt]{article}
\usepackage{latexsym}
\usepackage{amsmath}
\usepackage{amssymb}
\usepackage[dvips]{graphicx}
\usepackage{latexsym}
\usepackage{enumerate,graphicx}
\usepackage[toc,page]{appendix}
\usepackage{setspace}
\usepackage{color,soul}
\usepackage{multirow}

\newcommand{\CP}{\mathbb{CP}}

\newcommand{\R}{\mathbb{R}}

\newcommand{\Tr}{\mbox{Tr}}

\renewcommand{\d}{\mathrm{d}}

\newcommand{\koniec}{\begin{flushright}  $\Box $ \end{flushright}}
\def\be{\begin{equation}}
\def\ee{\end{equation}}

\def\vv{\varepsilon}

\def\Om{\Omega}

\def\g{\gamma}

\def\p{\partial}

\renewcommand{\d}{\mathrm{d}}

\def\mc{\mathcal}

\def\dsl{\displaystyle}

\def\bs{\boldsymbol}

\def\d{\dot}
\def\dd{\ddot}
\def\ddd{\dddot}

\def\a{\alpha}
\def\ll{\lambda}

\newtheorem{theo}{Theorem}[section] 
\newtheorem{prop}[theo]{Proposition}

\begin{document}
\title{
{\bf  A quadric ansatz method for a certain class of second order PDEs  \vskip 20pt}}

\vskip 100pt
\author{
Prim Plansangkate\thanks{
  Email: prim.p@psu.ac.th} \\[15pt]
{\sl Division of Computational Science, Faculty of Science,} \\
 {\sl Prince of Songkla University, Hat Yai, Songkhla, 90110 Thailand}.}

\date{}
\maketitle

\vskip 50pt

\begin{abstract}
We develop a procedure to implement the method of quadric ansatz to a class of second order partial differential equations (PDEs), which includes the four-dimensional K\"ahler-Einstein equation with symmetry and the one-sided type-D Einstein equation with nonzero scalar curvature.  The procedure, which reduces the PDEs to ordinary differential equations (ODEs), involves imposing additional constraints to the form of the ansatz, depending on the exact form of the PDEs.  Thus its applicability varies within the class.  We identify applicable subclasses of a class of modified Toda equations, one of which includes the K\"ahler-Einstein equation with symmetry, and give the corresponding reduced ODEs.  In addition, we obtain a quadric ansatz reduction for a family of Einstein-Weyl equations in arbitrary dimensions.   Lastly, we suggest a particular form of equations for which our procedure appears to be most effective as the aforementioned constraints vanish, and also present some reduction results using another type of ansatz, namely the hyperplane ansatz.
\end{abstract}

\newpage

\setcounter{equation}{0}

\section{Introduction}
It is known (e.g. \cite{T95, DT02, D03, F12}) that the method of quadric ansatz is applicable to partial differential equations (PDEs) of the form
\be \label{eqforquadric_comp}
\frac{\p}{\p x^i} \left(\, b^{ij}(u) \, \frac{\p u}{\p x^j}  \, \right) = 0,
\ee
where $x^i,$ $i=1, \dots, N,$ denote the $N$ independent variables, which will be written collectively as $\boldsymbol{x}=(x^i),$  ${\bf b}(u)$ is an $N \times N$ symmetric matrix whose components are functions of $u,$  and we use the summation convention over the repeated dummy indices.   
The quadric ansatz is an assumption of the existence of a solution  $u(\bs{x})$ given implicitly by
\be \label{quadricansatz}
M_{ij}(u) x^i x^j = C,
\ee
where ${\bf M}(u)$ is an $N \times N$ symmetric matrix whose components are functions of  $u,$ and $C$ is a constant.\footnote{The name quadric ansatz comes from the fact that a level set of the solution $u(\bs{x})$ gives a quadric hypersurface by (\ref{quadricansatz}).}   The method of quadric ansatz makes use of the ansatz to reduce the PDE to an ordinary differential equation (ODE) for  ${\bf M}(u),$ which would in principle be easier to solve. The method works well for equations of the form (\ref{eqforquadric_comp}), see for example \cite{T95, DT02, D03, F12}.

To our knowledge, the only use of the quadric ansatz to an equation that is not of the form (\ref{eqforquadric_comp}) was mentioned in \cite{T95}, where the ansatz was applied to the equation for four-dimensional K\"ahler-Einstein metrics with symmetry \cite{PP91}, 
\be \label{modTodaT95} u_{xx} + u_{yy} + (e^u)_{zz} - \frac{2}{z} (e^u)_z = 0,\ee
which is of the form (\ref{eqforquadric_comp}) plus an additional term involving the first order derivatives of $u$.  The result was then given in the case where the matrix ${\bf M}$ is assumed to be diagonal. 

Another equation which motivates our interest in exploring the applicability of the method of quadric ansatz to a wider class of equations than (\ref{eqforquadric_comp}) is the equation for one-sided type-D Einstein metrics with nonzero scalar curvature, excluding K\"ahler-Einstein metrics \cite{T21}, which is given by
\be \label{modTodaT21} u_{xx} + u_{yy} + (e^u)_{zz} + e^u \, \Big(  \frac{72 \Lambda z^2}{1- 12 \Lambda z^3}\, u_z - \frac{144 \Lambda z}{1- 12 \Lambda z^3} \Big) = 0,\ee
where $\Lambda$ is a nonzero constant.  The class of metrics is a generalisation of one-sided type-D vacuum metrics determined by the $SU(\infty)$-Toda equation \cite{PB84, PB87}, which includes the Chen-Teo instantons \cite{CT11}.

Both (\ref{modTodaT95}) and (\ref{modTodaT21}) are of the form
\[ \frac{\p}{\p x^i} \left(\, b^{ij}(u) \, \frac{\p u}{\p x^j}  \, \right) + T = 0, \qquad \mbox{where}\]
\be \label{modTodaT} \frac{\p}{\p x^i} \left(\, b^{ij}(u) \, \frac{\p u}{\p x^j}  \, \right)  =  u_{xx} + u_{yy} + (e^u)_{zz} \quad \mbox{and} \quad  T\,= \,e^u u_z \, f(z)  + e^u h(z)\ee
for some functions $f(z)$ and $h(z),$  where we note that the additional term $T$ only involves upto the first order derivatives of $u.$

This paper aims to explore in a general setting how the quadric ansatz may work for equations of the form 
\be \label{modeqforquadric}\frac{\p}{\p x^i} \left(\, b^{ij}(u) \, \frac{\p u}{\p x^j}  \, \right) + T(\boldsymbol{x}, u, u_{x^i}) = 0,\ee
where $T(\boldsymbol{x}, u, u_{x^i})$ is an additional term which may depend on the independent variables $\boldsymbol{x}=(x^i),$ the dependent variable $u$ and the first order derivatives ${\dsl u_{x^i} := \frac{\p u}{\p x^i}}.$  \, We find a procedure such that one can perform the quadric ansatz reduction in a similar fashion to what one would do to (\ref{eqforquadric_comp}), subject to a constraint on the matrix ${\bf M}$ of the ansatz (\ref{quadricansatz}), imposed by the term $T.$  

The constraints makes the applicability of our procedure varies with the exact form of $T.$  In particular, for equation 
(\ref{modTodaT}) we find that if $h(z) =0,$ the constraint allows ${\bf M}$ to be invertible  if and only if $\dsl f(z) = \frac{k}{z}$ for some constant $k.$  (See Proposition \ref{prop_Todacase1}.)    

For nonzero $h(z)$ however, we find that the complicated form of the constraint makes a calculation in the general case intractable.  Nevertheless, in the special case where ${\bf M}$ is assumed to be diagonal and invertible,  
we obtain a constraint on entries of ${\bf M}$ in terms of $f(z)$ and $h(z),$ and write down the reduced ODE for one particular example, where $\dsl f(z) = \frac{k}{z}$ and $\dsl h(z) = \frac{l}{z^2}$ for some constants $k$ and $l.$ (See Proposition \ref{prop_Todacase2}.)  In addition, our constraint implies that equation (\ref{modTodaT21}), corresponding to $\dsl f(z) = \frac{72 \Lambda z^2}{1- 12 \Lambda z^3}$ and $\dsl h(z) = - \frac{144 \Lambda z}{1- 12 \Lambda z^3},$ does not admit a quadric ansatz solution if ${\bf M}$ is diagonal and invertible.   Whether (\ref{modTodaT21}) admits any quadric ansatz solution is still an open problem.

\vspace{0.2cm}

The paper is organised as follows.  In Section 2, we write down a procedure to extend the method of quadric ansatz to an equation of the form (\ref{modeqforquadric}), and apply it to the class of equations (\ref{modTodaT}) in Section 3.  Then in Section 4, we search for quadric ansatz solutions for a class of Einstein-Weyl equations, which is a generalisation of the Kadomtsev-Petviashvili (dKP) equation to higher dimensions, and give the reduced ODE in the special case where ${\bf M}$ is assumed to be of a certain simple form.  Lastly, in Section 5 we discuss a form of the additional term $T$ in (\ref{modeqforquadric}) that is deemed optimal for our procedure, and also investigate a reduction by another type of ansatz, namely the hyperplane ansatz.


\section{An Extension of the Method of Quadric Ansatz}

\setcounter{equation}{0}

Let us first briefly review the method of quadric ansatz for a PDE of the form (\ref{eqforquadric_comp}),
\[ 
\frac{\p}{\p x^i} \left(\, b^{ij}(u) \, \frac{\p u}{\p x^j}  \, \right) = 0.
\]
Implicitly differentiating the quadric ansatz (\ref{quadricansatz}),
$\dsl 
M_{ij}(u) x^i x^j = C,
$
implies that 
\be \label{u_xk} \frac{\p u}{\p x^k} = - \frac{2}{\dot Q} M_{kj} x^j, \ee
where  $Q(\boldsymbol{x}, u)  := M_{ij}(u) x^i x^j$ and 
$\dsl \dot Q := \frac{\p Q}{\p u} =  \dot M_{ij} \, x^i x^j.$  \, Then
substituting (\ref{u_xk}) into (\ref{eqforquadric_comp}) gives 
\be \label{matrixeq1} {\bf X}^T  \left(   \frac{d}{du} ({\bf MbM})  - \frac{1}{2} \Tr({\bf bM})  \dot{\bf M}   -  \frac{{\bf X}^T ({\bf MbM}) {\bf X}}{{\bf X}^T \dot{\bf M} {\bf X}} \ddot{\bf M}   \right) {\bf X} = 0, \ee
where $\dsl \dot{\bf M}:= \frac{d{\bf M}}{du},$ ${\bf X}^T  = (x^1 \, \dots \, x^N)$ is a row vector and the transpose of ${\bf X}.$

The vanishing of the matrix in the bracket in (\ref{matrixeq1}) gives an equation which is regarded as a reduction of (\ref{eqforquadric_comp}) by the quadric ansatz.  Note, however, that for this equation to be an ODE for ${\bf M}$ with respect to the independent variable $u,$ one requires that $\dsl \frac{{\bf X}^T ({\bf MbM}) {\bf X}}{{\bf X}^T \dot{\bf M} {\bf X}}$ is a function of $u$ only, with no explicit $\bs{x}$-dependence.  With no prior knowledge of a solution $u(\boldsymbol{x}),$ treating $u, x^i$ as independent variables, it follows that 
\[ g\dot {\bf M} = {\bf M b M}, \]
for some function $g(u).$  Differentiating the above equation and comparing it with the vanishing of the matrix in the bracket of (\ref{matrixeq1}), 
$\dsl \frac{d}{du} ({\bf MbM})  - \frac{1}{2} \Tr({\bf bM})  \dot{\bf M}   -  g \ddot{\bf M} = 0,$
 implies that  $\dsl \dot g := \frac{d g}{du} = \frac{1}{2} \mbox{Tr}({\bf bM}).$

Hence altogether one obtains the quadric ansatz reduction of (\ref{eqforquadric_comp}) as a system of ODEs
\be \label{quadred} g\dot {\bf M} = {\bf M b M},  \qquad    \dot g = \frac{1}{2} \mbox{Tr}({\bf bM}).\ee

Also, as shown in \cite{T95}, if ${\bf M}$ is invertible, by introducing ${\bf N} = - {\bf M}^{-1}$ the system 
(\ref{quadred}) becomes
\be \label{Neq} g \dot {\bf N} = {\bf b}  \qquad  g^2 \det {\bf N} =  \zeta, \ee
where $\zeta$ is an arbitrary nonzero constant.   \, For many cases, system (\ref{Neq}) turns out to be easier to solve than (\ref{quadred}).

\vspace{0.5cm}

Now consider a PDE of the form (\ref{modeqforquadric}),
\[\frac{\p}{\p x^i} \left(\, b^{ij}(u) \, \frac{\p u}{\p x^j}  \, \right) + T(\boldsymbol{x}, u, u_{x^i}) = 0.\]
 By substituting (\ref{u_xk}) into equation (\ref{modeqforquadric}), one obtains instead of (\ref{matrixeq1}) the equation
\be  \label{matrixeq2} {\bf X}^T  \left(   \frac{d}{du} ({\bf MbM})  - \frac{1}{2} \Tr({\bf bM})  \dot{\bf M}   - \frac{{\bf X}^T ({\bf MbM}) {\bf X}}{{\bf X}^T \dot{\bf M} {\bf X}}  \ddot{\bf M} + \frac{\dot Q T}{4} \,  \dot{\bf M}  \right) {\bf X} = 0. \ee
A way to proceed is to follow the same argument and require that both functions
 $\dsl \frac{{\bf X}^T ({\bf MbM}) {\bf X}}{{\bf X}^T \dot{\bf M} {\bf X}}$ and  $\dsl \dot Q T$ depend only on $u$ with no explicit $\boldsymbol{x}$-dependence.  The requirement on  $\dsl \frac{{\bf X}^T ({\bf MbM}) {\bf X}}{{\bf X}^T \dot{\bf M} {\bf X}}$ gives the first equation of (\ref{quadred}) as before, with $g(u) = \dsl \frac{{\bf X}^T ({\bf MbM}) {\bf X}}{{\bf X}^T \dot{\bf M} {\bf X}}.$  

Requiring that $\dsl \dot Q T$  is also a function of $u$ puts another constraint on ${\bf M}.$  
Let us write this as the existence of a function $\dsl \tau(u)$ such that $\dsl \tau(u) = \frac{\dot Q T}{4}.$ Differentiating the first equation of (\ref{quadred}) and comparing it with the vanishing of the matrix in the bracket of (\ref{matrixeq2}), 
${\dsl   \frac{d}{du} ({\bf MbM})  - \frac{1}{2} \Tr({\bf bM})  \dot{\bf M}   - g  \ddot{\bf M} + \tau  \dot{\bf M}  = 0,}$
this time yields  $\dsl \dot g  = \frac{1}{2} \mbox{Tr}({\bf bM}) - \tau.$   \, Therefore we arrive at a system of ODEs almost the same as  (\ref{quadred}).   As before, the assumption of an invertible ${\bf M}$ yields another form of the system in terms of ${\bf N} = - {\bf M}^{-1}.$  \, The result is given in the following proposition, using the notation as defined previously.

\begin{prop} \label{prop_modquadric}
Given a PDE of the form {\rm(\ref{modeqforquadric})},
\[\frac{\p}{\p x^i} \left(\, b^{ij}(u) \, \frac{\p u}{\p x^j}  \, \right) + T(\boldsymbol{x}, u, u_{x^i}) = 0,\]
a function $u(\boldsymbol{x})$ defined implicitly by {\rm(\ref{quadricansatz})}, ${\bf X}^T {\bf M}(u) {\bf X} = C,$ is a solution of {\rm(\ref{modeqforquadric})} if there exist functions $g(u)$ and $\tau(u)$ such that $({\bf M}, g, \tau)$
satisfy the system
\begin{eqnarray}
g\dot {\bf M} &=& {\bf M b M}, \label{modquadredM} \\
\dot g &=& \frac{1}{2} {\rm{Tr}}({\bf bM}) - \tau, \label{modquadredg} \\
 \tau &=& \frac{({\bf X}^T \dot {\bf M} {\bf X}) \, T}{4}. \label{modquadredtau}
\end{eqnarray}
If ${\bf M}$ is invertible and let ${\bf N} = - {\bf M}^{-1},$ the system {\rm(\ref{modquadredM}, \ref{modquadredg})} is equivalent to
\begin{eqnarray}
g \dot {\bf N} &=& {\bf b}, \label{modquadredN} \\
g^2 \, e^{2\int (\tau/g) \, du} \, \det {\bf N} &=& \zeta, \label{modquadredNg} 
\end{eqnarray}
where  $\zeta$ is an arbitrary nonzero constant.
\end{prop}

In Section $3$ and $4$ we apply the method of quadric ansatz described in Proposition \ref{prop_modquadric} to two classes of PDEs.  Let us note here that in solving the system (\ref{modquadredM}, \ref{modquadredg}, \ref{modquadredtau}), one could use (\ref{modquadredtau}) as a constraint on ${\bf M}$ and seek a form of ${\bf M}$ (e.g. setting some entries of ${\bf M}$ to zero) such that $({\bf X}^T \dot {\bf M} {\bf X}) \, T$ is explicitly independent of ${\bs x}.$  As (\ref{modquadredtau}) does not involve ${\bf b},$ this form of ${\bf M}$ will depend only on the additional term $T$ and is independent of the part 
$\dsl \frac{\p}{\p x^i} \left(\, b^{ij}(u) \, \frac{\p u}{\p x^j}  \, \right)$ of  (\ref{modeqforquadric}).  One can then assume this form of ${\bf M}$ to solve  (\ref{modquadredM}, \ref{modquadredg}), or (\ref{modquadredN}, \ref{modquadredNg}).  We find however that in this approach  the term $T$ may put a severe constraint on ${\bf M}$ such that imposing (\ref{modquadredM}, \ref{modquadredg}) results in ${\bf M}=0.$   While this happens, we also find that it is possible to find some other less restrictive form of ${\bf M}$ such that (\ref{modquadredtau}) holds by means of  (\ref{modquadredM}) and the quadric ansatz ${\bf X}^T {\bf M}(u) {\bf X} = C$ itself.  Such instances are demonstrated in Section 3.2 and 4.


\section{A class of modified Toda equations} 

\setcounter{equation}{0}

The $SU(\infty)$-Toda equation, 
\be \label{Todaeq}\dsl u_{xx} + u_{yy} + (e^u)_{zz}=0, \ee
 appears in many contexts in geometry.  For example, it determines scalar-flat K\"ahler metrics with symmetry \cite{LeBrun91} and one-sided type-D vacuum metrics \cite{PB84, PB87}.
The equation is of the form (\ref{eqforquadric_comp}), where 
\be \label{bforToda}  {\bf b}   = \left(
\begin{array}{ccc}
1 & 0 & 0 \\
0 & 1 & 0 \\
0 & 0 & e^u
\end{array}  \right). \ee
It was shown in \cite{T95} that  under the quadric ansatz assumption, (\ref{Todaeq}) reduces to a special case of the Painlev\'e III equation.  As mentioned in the Introduction, in the same article \cite{T95} the quadric ansatz was also applied to equation (\ref{modTodaT95})
\[ u_{xx} + u_{yy} + (e^u)_{zz} - \frac{2}{z} (e^u)_z = 0,\]
which determines four-dimensional K\"ahler-Einstein metrics with symmetry \cite{PP91}, and is not of the form (\ref{eqforquadric_comp}).  The result was stated in the case where the matrix ${\bf M}$ of the quadric ansatz is assumed to be diagonal, that the resulting ODE is equivalent to the system of ODEs found in \cite{DS94} and does not possess the Painlev\'e property.  

Motivated by this and the recent interest in equation (\ref{modTodaT21})
\[ u_{xx} + u_{yy} + (e^u)_{zz} + e^u \, \Big(  \frac{72 \Lambda z^2}{1- 12 \Lambda z^3}\, u_z - \frac{144 \Lambda z}{1- 12 \Lambda z^3} \Big) = 0,\]
which determines one-sided type-D Einstein metrics with nonzero scalar curvature, excluding K\"ahler–Einstein metrics \cite{T21}, 
in this section we apply the method of quadric ansatz described in Proposition \ref{prop_modquadric} to the modified Toda equation 
\be \label{modTodaeq} u_{xx} + u_{yy} + (e^u)_{zz} + e^u u_z \, f(z)  + e^u h(z) =0,\ee
where $f(z)$ and $h(z)$ are some functions.  In the notation of Proposition \ref{prop_modquadric}, this equation is of the form (\ref{modTodaT})  
\[ \frac{\p}{\p x^i} \left(\, b^{ij}(u) \, \frac{\p u}{\p x^j}  \, \right) + T = 0, \qquad \mbox{where}\]
\[ \frac{\p}{\p x^i} \left(\, b^{ij}(u) \, \frac{\p u}{\p x^j}  \, \right)  =  u_{xx} + u_{yy} + (e^u)_{zz} \quad \mbox{and} \quad  T\,= \,e^u u_z \, f(z)  + e^u h(z).\]

\vspace{0.3cm}

Below we discuss possible quadric ansatz reductions, under the assumption that the matrix ${\bf M}$ of the quadric ansatz is invertible.  

\vspace{0.5cm}

Let ${\bf N} = - {\bf M}^{-1}.$  Then (\ref{modquadredN}) implies that \cite{T95}
 \be \label{NToda} {\bf N}   = \left(
\begin{array}{ccc}
X & \a & \beta \\
\a & Y & \g \\
\beta  & \g & Z
\end{array}  \right),\ee
where  $\a, \beta, \g$ are constants and the functions $X,Y,Z$ of $u$ satisfy
\be \label{modTXYZeq} \dot X = \dot Y = \frac{1}{g}, \qquad \dot Z = \frac{e^u}{g}. \ee
In turn, (\ref{modTXYZeq}) implies that 
\be \label{Todaeta} Y = X + \eta, \quad \mbox{for some constant} \; \eta, \quad \mbox{and} \quad \dot Z =  e^{u} \dot X.\ee

\subsection{Case 1:  $\bs{h(z)=0}$} \label{sec:Todacase1}

If $h(z)=0,$ then $T\,= \,e^u u_z \, f(z)$ and, using (\ref{u_xk}), 
condition (\ref{modquadredtau}) becomes
\[\tau(u) =  -\frac{1}{2} e^u f(z) \,M_{3j}(u)x^j,\]
where the summation is now over the repeated dummy index which runs over $\{1,2,3\},$ with $x^1 =x,$\, $x^2 = y,$ \,$x^3 =z.$   For the right-hand side of the above equation to have no explicit $\boldsymbol{x}$-dependence, it follows that $M_{31}=M_{32}=0.$  Also, under the assumption that ${\bf M}$ is invertible, which means $M_{33}$ cannot vanish, $f(z)$ must be of the form $\dsl f(z) = \frac{k}{z}$ for some constant $k.$  Under these constraints, we analyse the quadric ansatz reduction and give the outcome in Proposition \ref{prop_Todacase1} below.

\begin{prop} \label{prop_Todacase1}
An equation of the form
\[u_{xx} + u_{yy} + (e^u)_{zz}+e^u f(z) \, u_z = 0\]
admits solutions given implicitly by the quadric ansatz {\rm (\ref{quadricansatz})},  ${\bf X}^T {\bf M}(u) {\bf X} = C,$ where ${\bf M}$ is invertible, if and only if $\dsl f(z) = \frac{k}{z}$  for some constant $k$  and 
\be \label{Ncase11} {\bf N} := - {\bf M}^{-1} =  \left(
\begin{array}{ccc}
X & \a & 0 \\
\a & Y & 0 \\
0  & 0 & Z
\end{array}  \right).\ee
Then the quadric ansatz reduction is given by an ODE for $X,$ which can be divided into two cases {\rm:}
\begin{enumerate}[$1.$]
\item $k=-1:$ \, The ODE is of first order and solvable by elementary functions.

\item $k\ne -1:$ \, The ODE is of second order.  Letting $v=e^u,$ if the ODE in the $v$ variable is rational in $X$ and $\dsl \frac{dX}{dv},$ then it possesses the Painlev\'e property if and only if $k=0,$ which corresponds to the $SU(\infty)$-Toda equation {\rm(\ref{Todaeq})}. 
\end{enumerate}

\end{prop}

\vspace{0.5cm}

\noindent {\bf Proof.} \;  The proof is by direct calculation.  The first part of the proposition comes from imposing  (\ref{modquadredtau}) as discussed in the paragraph before the proposition.  

Then with $\dsl \tau = \frac{k e^u}{2Z},$ (\ref{modquadredNg})  becomes
\[ g^2 Z^k \det{\bf N} = \zeta,  \]
where we have absorbed a constant of integration into $ \zeta.$ \,  Now, the above equation together with  (\ref{modTXYZeq})  yields
\be \label{eq1NmodT1.1XZ} \dot Z =  e^{u} \dot X,   \qquad  Z^{k+1} (X^2 + \eta X -\a^2) = \zeta {\dot X}^2,\ee
which can be divided into two cases.

If $k=-1,$ the second equation of (\ref{eq1NmodT1.1XZ}) is a first order ODE for $X$ and solvable in terms of elementary functions.  

If $k\ne -1,$ system (\ref{eq1NmodT1.1XZ}) gives a second order ODE for $X,$ 
which upon changing the independent variable to $v=e^u,$ is given by
\be \label{2ndredeqmodT1.1v}
X'' - \frac{2X+\eta}{2(X^2 +\eta X -\a^2)} (X')^2 + \frac{1}{v} X' 
- \frac{k+1}{2 \zeta^\frac{1}{k+1}} v^{\frac{k-1}{k+1}} (X^2 +\eta X -\a^2)^\frac{1}{k+1}  (X')^\frac{2k}{k+1} = 0,
\ee
where $\dsl X' = \frac{dX}{dv},$ etc.

When $k=0,$  equation
\be \label{modTcase1eq} u_{xx} + u_{yy} + (e^u)_{zz}+ \frac{k}{z} e^u \, u_z = 0 \ee
becomes the $SU(\infty)$-Toda equation (\ref{Todaeq}).  One can use a M\"obius transformation of $X$ to put  (\ref{2ndredeqmodT1.1v}) into the Painlev\'e V equation with special values of parameters such that it can be further transformed \cite{FA82} to the Painlev\'e III equation.  This result is already known in \cite{T95}. 

Now assume that the values of $k\in \R$ is such that  (\ref{2ndredeqmodT1.1v}) is of the form
\be \label{rational2nd} X'' = F(X', X, v),\ee
where $F$ is rational in $X'$ and $X,$ and analytic in $v.$ \,  Within the class (\ref{rational2nd}), it is well known (see e.g. \cite{Ince}) that an equation of Painlev\'e type is necessarily of the form
\be \label{necC1}X'' = A(X,v)(X')^2 + B(X,v) X' + C(X,v),\ee
where $A, B, C$ are rational functions of $X$ and with coefficients analytic in $v.$    Then it can be seen that for generic values of $\a, \eta$ in (\ref{2ndredeqmodT1.1v}), only $k=0$ satisfies the necessary condition (\ref{necC1}).  Note that in the special case where $X^2 +\eta X -\a^2$ is a perfect square, both $k=0, 1$ satisfy (\ref{necC1}).  However, it can be shown that the ODE (\ref{2ndredeqmodT1.1v}) with $k=1$ fails the test \cite{ARS80} for Painlev\'e property, by an appearance of a logarithmic singularity.   \koniec

\noindent {\bf Remarks.}

\noindent 1. For $k=-1,$ the second equation  of (\ref{eq1NmodT1.1XZ}) becomes
$\dsl \zeta {\dot X}^2 = X^2 + \eta X -\a^2,$
which can be solved to give
\[X = \frac{1}{2} \left( \kappa e^{\frac{\pm u}{\sqrt{\zeta}}} - \eta + \frac{\a^2 + \dsl \frac{\eta^2}{4}}{\kappa e^{\frac{\pm u}{\sqrt{\zeta}}}}\right).\]
Then $Y$ and $Z$ are given by (\ref{Todaeta}), and the quadric ansatz solution can be worked out from (\ref{Ncase11}).

For example, setting $\zeta =1$ and $\a = \eta =0,$  the quadric ansatz solution to \\ $\dsl  u_{xx} + u_{yy} + (e^u)_{zz} - \frac{e^u}{z}  \, u_z = 0$ can be inverted to given an explicit solution $u(x,y,z)$ given by
\[ u = \ln\left(\frac{x^2+y^2+\sqrt{(x^2+y^2)^2 + 4 c \, z^2}}{c}\right), \]
where $c$ is an arbitrary nonzero constant.

\vspace{0.5cm}

\noindent 2.  One can also look for a quadric ansatz solution in the case where ${\bf M}$ is singular.   The condition (\ref{modquadredtau}) constrains ${\bf M}$ to be of the form
\[  {\bf M}   = \left(
\begin{array}{ccc}
{\mc X} & {\mc A} & 0 \\
{\mc A} & {\mc Y} & 0 \\
0 & 0 & {\mc Z}
\end{array}  \right),\]
where ${\mc X}, {\mc Y}, {\mc Z}, {\mc A}$ are functions of $u.$

If  ${\mc Z} = 0,$ the quadric ansatz (\ref{quadricansatz}) implicitly defines a function $u$ which depends only on $x$ and $y$ variables, and it effectively becomes the  ansatz for the two-dimensional Laplace's equation.  See \cite{D03} for quadric ansatz solutions to  Laplace's equation in arbitrary dimensions.

If ${\mc Z} \ne0,$ it follows that ${\mc X}{\mc Y} = {\mc A}^2.$
The resulting ODE is again solvable in the case $k=-1,$ where the implicit solution to $\dsl  u_{xx} + u_{yy} + (e^u)_{zz} - \frac{e^u}{z}  \, u_z = 0$ is given by
\[(\a y+ x)^2 {\mc X} + z^2 {\mc Z} = C, \quad \mbox{where} \quad {\mc X} = \frac{1}{(c_1 u +c_2)^2}, \; {\mc Z}= \frac{(1+\a^2)}{2c_1 e^u(c_1(u-1) +c_2) + c_3},\]
and $C, c_1, c_2, c_3$ are arbitrary constants. 
The case $k\ne -1$ gives a third order ODE for ${\mc X}$ with no explicit term on the independent variable $u.$  

\vspace{0.5cm}

\noindent 3.  It is stated in \cite{T21}, with reference to \cite{BFKN22}, that the only modifications of the $SU(\infty)$-Toda equation of the form (\ref{modTodaeq}),
\[u_{xx} + u_{yy} + (e^u)_{zz} + e^u u_z \, f(z)  + e^u h(z) =0,\]
that are integrable are those related to the $SU(\infty)$-Toda equation (\ref{Todaeq}) by the transformation
\[z \to Z = \rho(z), \qquad u \to U = u+\Om(z),\]
for some functions $\rho(z)$ and $\Om(z).$

It can be shown by direct calculation that  (\ref{modTodaeq}) satisfies this integrability condition if and only if
\[\frac{f^2}{9} +\frac{f'}{3} - \frac{h}{2} =0.\]
Then it follows that (\ref{modTcase1eq}) is integrable if and only if $k=0, 3.$

While the integrability of the $SU(\infty)$-Toda equation ($k=0$) can be detected from the quadric ansatz reduction which yields the Painlev\'e III equation,  the case $k=3$ gives the ODE (\ref{2ndredeqmodT1.1v}) that is not rational in $X$ and $X'.$  Its Painlev\'e analysis, which may require further transformation of variables, is left for a future work.

\vspace{0.5cm}

\subsection{Case 2:  $\bs{h(z) \ne 0}$} \label{sec_Todacase2}

With $T\,= \,e^u u_z \, f(z)  + e^u h(z),$  Proposition \ref{prop_modquadric} requires an existence of a function $\tau(u)$ such that (\ref{modquadredtau}), 
\be  \label{modTodacase2tau} \tau(u) =  -\frac{1}{2} e^u f(z) \,M_{3j}(u)x^j + \frac{1}{4}  e^u h(z) \, \dot M_{rs}(u) x^r x^s,\ee
where $j,r,s = 1,2,3$ and $x^1 =x,$\, $x^2 = y,$ \,$x^3 =z,$ is satisfied.

We find that it is simpler to find a set of constraints on ${\bf M},$ such that the right-hand side of (\ref{modTodacase2tau}) does not depend explicitly on $\bs{x},$ if one does not make use of (\ref{modquadredM}), although this form of ${\bf M}$ will be more restrictive than necessary and only give a subclass of solutions. It turns out however that for the modified Toda equation (\ref{modTodaeq}) this subclass is not interesting, as it only consists of solutions independent of the $x, y$ variables and is only compatible with certain forms of $f(z)$ and $h(z).$  (See the remark after Proposition \ref{prop_Todacase2}.) 

If one makes use of (\ref{modquadredM}), the form of ${\bf M}$ will be less restrictive, but we find that a calculation for the  general case is intractable due to the complicated form of (\ref{modTodacase2tau}).  
Nevertheless, in Proposition \ref{prop_Todacase2} below we present a result where the matrix ${\bf M}$ of the quadric ansatz is assumed to be diagonal and invertible.

\begin{prop} \label{prop_Todacase2}
An equation of the form {\rm(\ref{modTodaeq})},
\[u_{xx} + u_{yy} + (e^u)_{zz}+e^u f(z) \, u_z + e^u h(z) = 0,\]
admits solutions given implicitly by the quadric ansatz {\rm (\ref{quadricansatz})}, ${\bf X}^T {\bf M}(u) {\bf X} = C,$ where ${\bf M}$ is diagonal and invertible, if and only if ${\bf N} = - {\bf M}^{-1}$ is of the form
\[ {\bf N} = \left(
\begin{array}{ccc}
X & 0 & 0 \\
0 & X & 0 \\
0  & 0 & Z
\end{array}  \right)\]
and the entries of ${\bf N},$ $f(z)$ and $h(z)$ satisfy
\be \label{Todacase2fhsubprop}2 (zf'+f) +(z^2h'+2zh) \frac{d}{du}\ln\left(\frac{Z}{X}\right) - C h' \frac{Z \dot X}{X} =0.\ee

If the constant $C=0,$ $\dsl f(z) = \frac{k}{z},$ $\dsl h(z)=\frac{l}{z^2}$ for some constants $k$ and $l,$
then the quadric ansatz reduction can be written as a second order ODE for $W(Z) := \dot Z(Z(u)),$ given by
 \begin{eqnarray}& & (4Z-lW) \, W''  + 2l \, (W')^2 - \left( 4(k+1) +3l +\frac{l}{2}(k+9)\frac{W}{Z} \right) W' \qquad  \nonumber \\
& & + \; \left(\frac{k+5}{2}\, l +(k+1)(k+3)\right) \frac{W}{Z} + l\left(k + \frac{l}{4}+3\right) \frac{W^2}{Z^2} + 2(k+1)+l \; = \; 0, \qquad \qquad \label{Todacase2Weq}
\end{eqnarray}
subject to
\[W' -\frac{l}{4} \, \frac{W^2}{Z^2} - \frac{k+1}{2} \, \frac{W}{Z} -1 \; \ne \; 0, \]
where $\dsl W' := \frac{dW}{dZ}.$
\end{prop}

\vspace{0.5cm}

\noindent {\bf Proof.} \;  Assume that ${\bf M}$ is diagonal and invertible, let
\[ {\bf N} = - {\bf M}^{-1}   = \left(
\begin{array}{ccc}
X & 0 & 0 \\
0 & Y & 0 \\
0  & 0 & Z
\end{array}  \right).\]
Equation (\ref{modquadredN}) with ${\bf b}$ given by  (\ref{bforToda}) implies that $X,Y,Z$ satisfy
(\ref{modTXYZeq}).  Then (\ref{modTodacase2tau}) becomes
\be \label{tauproof3.2} \tau(u) =  \frac{1}{2} e^u f(z) \,\frac{z}{Z} + \frac{1}{4}  e^u h(z) \, \dot X \left( \frac{x^2}{X^2} + \frac{y^2}{Y^2} +\frac{e^u z^2}{Z^2} \right).\ee
Now, from (\ref{Todaeta}) we have $Y = X+\eta.$ One finds that, using the quadric ansatz (\ref{quadricansatz}),
\[-\frac{x^2}{X} - \frac{y^2}{X+\eta} -\frac{z^2}{Z} = C,\]
the right-hand side of (\ref{tauproof3.2}) is explicitly independent of $x$ and $y$ if and only if $\eta=0.$  We now have
\[ \tau(u) =  \frac{1}{2} e^u f(z) \,\frac{z}{Z} + \frac{1}{4}  e^u h(z) \, \dot X \left( -\frac{1}{X} \Big(C +\frac{z^2}{Z} \Big) \frac{e^u z^2}{Z^2} \right).\]
Then differentiating both sides of the equation with respect to $z,$ holding $u$ constant, gives (\ref{Todacase2fhsubprop}).

If $C=0,$ $\dsl f(z) = \frac{k}{z}$ and $\dsl h(z)=\frac{l}{z^2}$ for some constants $k, l,$  equation (\ref{Todacase2fhsubprop}) is satisfied and
(\ref{tauproof3.2}) becomes
\[\tau = \frac{e^u }{2 Z} \left( k + \frac{l}{2} \left( \frac{e^u}{Z} - \frac{1}{X} \right)  \dot X  \right).\]
Now, to obtain the quadric ansatz reduction, one can use (\ref{modquadredg}) instead of (\ref{modquadredNg}), in which 
 the integral $\dsl \int \frac{\tau}{g} \, du$
cannot be evaluated to get a term in $X, Z$ and their derivatives.  Differentiating $\dsl g = \frac{e^u}{\dot Z}$ from (\ref{modTXYZeq}) and 
equating it to the right-hand side of (\ref{modquadredg}) gives an ODE which can be algebraically solved for $X$ in terms of $Z, \dot Z, \ddot Z.$  Then, differentiating $X$ once more and using (\ref{modTXYZeq}) yields a third order ODE for $Z.$  The order can be reduced by the change of variable  $W(Z) := \dot Z(Z(u)),$ which results in (\ref{Todacase2Weq}).
 \koniec

\noindent {\bf Remark.}  \, Consider an equation of the form 
\be \label{geneqTT21} \frac{\p}{\p x^i} \left(\, b^{ij}(u) \, \frac{\p u}{\p x^j}  \, \right) + T =0, \qquad \mbox{with} \quad T\,= \,e^u u_z \, f(z)  + e^u h(z),\ee
where $i, j = 1,2,3$ and $x^1 =x,$\, $x^2 = y,$ \,$x^3 =z,$ with an arbitrary matrix ${\bf b}(u) = \big( b^{ij}(u) \big).$

A subclass of quadric ansatz solutions of (\ref{geneqTT21}) consists of solutions coming from
 ${\bf M}$ such that $({\bf X}^T \dot {\bf M} {\bf X}) \, T$ is explicitly independent of $\bs{x},$ regardless of the matrix ${\bf b}$ defining (\ref{geneqTT21}), i.e. ${\bf M}$ is of the form that 
 (\ref{modquadredtau}) holds without assuming (\ref{modquadredM}).

For the modified Toda equation (\ref{modTodaT}), it turns out however that this subclass of solutions is not interesting, as 
the solutions are independent of the $x$ and $y$ variables, i.e. the quadric  ansatz reduces to an ansatz $\dsl {\mc Z}(u) = \frac{C}{z^2}$ to the ODE ${u_{zz}+u_z^2+ f(z) \, u_z +  h(z) = 0.}$

\vspace{0.3cm}

Finally, let us note that for equation (\ref{modTodaT21}) of  \cite{T21}, the functions  
$\dsl f(z) = \frac{72 \Lambda z^2}{1- 12 \Lambda z^3}$ and $\dsl h(z) = - \frac{144 \Lambda z}{1- 12 \Lambda z^3}$
satisfy (\ref{Todacase2fhsubprop}) if and only if $\dsl \frac{ C Z \dot X}{X} =0$ and $\dsl  \frac{d}{du}\ln\left(\frac{Z}{X}\right) =1.$  This leads to a particular form of ${\bf M}$ which can be shown to be inconsistent with (\ref{modTodaT21}).  Thus we conclude that (\ref{modTodaT21})  does not admit a quadric ansatz solution with ${\bf M}$ diagonal and invertible.


\section{The $n$-dKP Einstein-Weyl equation}

\setcounter{equation}{0}

Recall that an Einstein-Weyl (EW) structure is a triplet $(\mc{W}, [h], D)$ consisting of a manifold
$\mc{W},$ a conformal class of metrics $[h]$ and a torsion-free connection $D$ such that $D$ is compatible with 
$[h]$ in the sense that, choosing a representative metric $h \in [h],$ $D h = \nu \otimes h$ for some one-form $\nu,$ and 
the symmetrised Ricci tensor of $D$ is proportional to $h.$  Locally, the structure is determined by a system of PDEs for the components of $h$ and $\nu.$
\, In many instances, the system of PDEs reduces to a single scalar PDE. \, For example , the dispersionless Kadomtsev-Petviashvili (dKP) equation,
\be \label{dKPeq} u_{xt} - (u u_x)_x - u_{y y}  =0,\ee
is well known to determine a $3$-dimensional Lorentzian EW structure admitting a parallel weighted vector field \cite{DMT01}.

The dKP equation and one of its generalisation, namely the $mn$-dKP equation \cite{SS16}, 
\[u_{xt} - (u^m u_x)_x - u_{y_1 y_1} - \dots - u_{y_n y_n} =0,\]
are of the form (\ref{eqforquadric_comp}).  Their quadric ansatz solutions have been studied in \cite{DT02, DP22}.  

Another generalisation of the  dKP equation to $n+2$ dimensions is given by
\be \label{dKPEWeq}  
u_{xt} - (u u_x)_x - u_{y_1 y_1} - \dots - u_{y_n y_n}  + \frac{2(n-1)}{n} u_x^2 =0,
\ee
where $n \ge 1$ is an integer.  It is known \cite{MOP12, DG21} to govern  a class of $(n+2)$-dimensional EW structures admitting a parallel null weighted vector field with a further assumption on the holonomy of the Weyl connection.
Equation (\ref{dKPEWeq}), which we shall call the $n$-dKP Einstein-Weyl ($n$-dKP EW) equation, is however not of the form (\ref{eqforquadric_comp}).  It is of the form (\ref{modeqforquadric}), 
$\dsl  \frac{\p}{\p x^i} \left(\, b^{ij}(u) \, \frac{\p u}{\p x^j}  \, \right) + T = 0,$ where
$\dsl \frac{\p}{\p x^i} \left(\, b^{ij}(u) \, \frac{\p u}{\p x^j}  \, \right)  =  u_{xt} - (u u_x)_x - u_{y_1 y_1} - \dots - u_{y_n y_n}$  and $\dsl T\,= \, \ll u_x^2,$  with $\dsl \ll =  \frac{2(n-1)}{n}.$

\vspace{0.5cm}

 In this section, we apply the method of the quadric ansatz described in Proposition \ref{prop_modquadric} to (\ref{dKPEWeq}).  Our result is similar to that of Section \ref{sec_Todacase2}.  That is, the constraints from (\ref{modquadredtau}) turn out to be too restrictive if one seeks a form of ${\bf M}$ that satisfies (\ref{modquadredtau}) without assuming (\ref{modquadredM}).  Nevertheless, using (\ref{modquadredM}) and the quadric ansatz itself, it is possible to find a nontrivial reduction if ${\bf M}$ is assumed to be invertible and of a certain simple form.  The result is summarised in Proposition \ref{prop_ndKPEW} below.

\begin{prop} \label{prop_ndKPEW}
Equation {\rm(\ref{dKPEWeq})},
\[u_{xt} - (u u_x)_x - u_{y_1 y_1} - \dots - u_{y_n y_n}  + \frac{2(n-1)}{n} u_x^2 =0,\]
admits solutions given implicitly by the quadric ansatz {\rm (\ref{quadricansatz})}, ${\bf X}^T {\bf M}(u) {\bf X} = C,$ where ${\bf M}$ is invertible and such that ${\bf N} = - {\bf M}^{-1}$ is of the form
\[ {\bf N} =   \left(\begin{array}{c|ccc|c}
   Y & 0 & \cdots &  0 &  \; Z  \\ \hline
   0  &   & & &   \; 0  \\
\vdots & &{\scalebox{1.2}{${\bf X}$}} &  & \; \vdots \\
   0  & & & &  \; 0   \\ \hline
Z & 0  & \cdots & 0  & \;  0 
  \end{array}\right),
\]
if and only if  \, $\dsl {\bf X}(u)=-2Z(u){\bf 1}_n$ and $C=0.$

Assume $n>1,$ the quadric ansatz reduction gives a third order ODE for $Z(u)$ given by
\be \label{ndKPEWredprop} \frac{-8(n-1) Z^2 \d Z \, \ddd Z + 12 \left(n - \frac{2}{3}\right) Z^2 (\dd Z)^2 + 4 (n-4) Z (\d Z)^2 \dd Z + (n+2)(n-2)(n-4)(\d Z)^4}{\big( (n+2)(\d Z)^2 -2 Z \dd Z \big)^2}
 = 0.\ee
\end{prop}

\vspace{0.5cm}

\noindent {\bf Proof.} \;  The $n$-dKP EW equation (\ref{dKPEWeq}) can be written in the form 
(\ref{modeqforquadric}), where
\[    {\bf b}(u)   =
  \left(\begin{array}{c|ccc|c}
   -u & 0 & \cdots &  0 &  \; \frac{1}{2} \\ \hline
    0  &   & & &   \;  0 \\
\vdots & &{\scalebox{1.2}{$-{\bf 1}_n$}} &  & \; \vdots \\
    0 & & & &  \; 0  \\ \hline
\frac{1}{2} & 0 & \cdots &  0 & \;  0
  \end{array}\right), \quad \mbox{and} \quad T\,= \, \ll u_x^2 \quad \mbox{with} \quad \ll =  \frac{2(n-1)}{n}.
\]

Recall from \cite{DP22} that if ${\bf M}$ is invertible, then equation (\ref{modquadredN}), $g  \dot {\bf N} = {\bf b},$ implies that 
\be \label{matrixN}   {\bf N}   =  - {\bf M}^{-1} =  
  \left(\begin{array}{c|ccc|c}
   Y & \beta_1 & \cdots &  \beta_n &  \; Z  \\ \hline
   \beta_1  &   & & &   \;  \vv_1 \\
\vdots & &{\scalebox{1.2}{${\bf X}$}} &  & \; \vdots \\
    \beta_n & & & &  \; \vv_n  \\ \hline
Z & \vv_1 & \cdots & \vv_n & \;  \phi
  \end{array}\right),
\ee
where $\beta_i, \vv_i,$ $i = 1,\dots, n,$ and $\phi$ are constants,  $Y$ and $Z$ are functions of $u,$  and ${\bf X}$ is an $n \times n$ symmetric matrix,
whose diagonal components $X_1, X_2, \dots, X_n$
are functions of $u$ and off-diagonal components $\alpha_{ij}, i>j$ are all constants.  
Equation (\ref{modquadredN}) also implies that 
\be \label{dKPEWbeq}
\dot Y = -\frac{u}{g},\quad \dot {X_i} = -\frac{1}{g}, \quad  \dot Z = \frac{1}{2g}.
\ee
The system (\ref{dKPEWbeq}) gives
\[ 
\dot Y + 2 u \dot Z  = 0, 
\quad
\mbox{and}
\quad  X_i = -2 Z + \g_i, \quad \mbox{for all} \quad i = 1, \dots, n,
\]
where $\gamma_i$ are constants.

Now consider only a simple case where all constants $\beta_i, \vv_i,$ $i = 1,\dots, n,$ and $\phi$ are zero, that is
\[  {\bf N}   =
  \left(\begin{array}{c|ccc|c}
   Y & 0 & \cdots &  0 &  \; Z  \\ \hline
   0  &   & & &   \; 0  \\
\vdots & &{\scalebox{1.2}{${\bf X}$}} &  & \; \vdots \\
   0  & & & &  \; 0   \\ \hline
Z & 0  & \cdots & 0  & \;  0
  \end{array}\right). \]

Equation (\ref{modquadredtau}) becomes $\dsl \tau(u) = \frac{\ll t^2}{\dot Q Z^2},$ which is satisfied if and only if $\dot Q$ is of the form $\dot Q = \rho(u) \, t^2$ for some function $ \rho(u).$  \, Using the quadric ansatz  (\ref{quadricansatz}),
\[ - 2\frac{xt}{Z} - \frac{y_1^2}{\g_1 -2Z} + \dots + \frac{y_n^2}{\g_n -2Z}  + \frac{Y t^2}{Z^2} =C,\]
it follows that 
\[\dot Q=   -\frac{\dot Z}{Z} \, \left(C + \frac{\g_1 y_1^2}{(\g_1 -2Z)^2} + \dots + \frac{\g_n y_n^2}{(\g_n -2Z)^2} - \frac{Y t^2}{Z^2}\right) - \frac{2\dot Z}{Z^2} \, \left(u+\frac{Y}{Z}\right) t^2,\] 
which is of the required form if and only if $\g_i =0,$ i.e. $X_i = -2Z,$ for all $i = 1, \dots, n$ and $C=0.$

Then (\ref{modquadredtau}) holds with
\[\tau = - \ll \, \frac{Z}{\dot Z (2uZ+Y)}.\]
Next, as the integral $\dsl \int \frac{\tau}{g} \, du$ in (\ref{modquadredNg}) cannot be simply evaluated to give a term in $Y, Z$ and their derivatives, we proceed to use (\ref{modquadredg}), $ \dot g = \frac{1}{2} \rm{Tr}({\bf bM}) - \tau,$ instead.  
This gives 
\be \label{ndKPEWred1} \frac{\ddot Z}{2 \dot Z^2} = \frac{n+2}{4Z} - \frac{2(n-1)}{n} \frac{Z}{\dot Z (2uZ+Y)},\ee
where we have put back $\dsl \ll = \frac{2(n-1)}{n}.$

If $n=1,$ then the $n$-dKP EW equation (\ref{dKPEWeq}) reduces to the dKP equation whose quadric ansatz solutions have been classified in \cite{DT02}.

For $n >1,$  (\ref{ndKPEWred1}) can be solved algebraically to get an expression for $Y$ in terms of $Z$ and its derivatives.  This, together with (\ref{dKPEWbeq}), give (\ref{ndKPEWredprop}).
\koniec

\noindent {\bf Example $\bs{4.1}$} \, Note that the numerator of (\ref{ndKPEWredprop}) is a third order ODE which has no explicit term in $u.$  The order can thus be reduced by changing the variable to ${W(Z) = \d Z(Z(u)).}$  This gives
\be \label{ndKPEWredWeq}  W'' - \frac{n}{2(n-1)} \frac{(W')^2}{W} - \frac{n-4}{2(n-1)} \frac{W'}{Z} - \frac{(n+2)(n-2)(n-4)}{8(n-1)} \frac{W}{Z^2} =0.\ee
One can find a solution of (\ref{ndKPEWredprop}) by solving (\ref{ndKPEWredWeq}) and integrating $\d Z = W(Z),$  provided that the denominator of (\ref{ndKPEWredprop}) is not zero.

For $n=2,$ (\ref{ndKPEWredWeq}) becomes
\[W'' -  \frac{(W')^2}{W}+\frac{W'}{Z} = 0, \]
which is the Painlev\'e III equation with zero parameters.  A solution of this leads to 
\[Z(u) = (a u + b)^m,\]
where $a, b$ are arbitrary constants and $m \ne 0, -1$ is another constant.  The quadric ansatz
\[ -2\frac{xt}{Z} - \frac{y_1^2 +y_2^2}{X}  + \frac{Y t^2}{Z^2} = 0\]
then gives a simple solution to the $n$-dKP EW equation (\ref{dKPEWeq}), which can be inverted to an explicit solution
\be \label{udKPEWsimpex} u = \a \, \frac{y_1^2 +y_2^2 - 4xt}{t^2} +\beta,\ee
where $\a \ne 0$ and $\beta$ are arbitrary constants. 

For $n>2,$ it can be shown that (\ref{dKPEWeq}) admits a solution of the form (\ref{udKPEWsimpex}), but necessarily with $\dsl \a = \frac{n}{8}$ while  $\beta$ remains arbitrary.

Let us note however that these simple solutions can be easily obtained by some other methods such as the method of symmetry.  We leave the analysis of other solutions for $n \ge 2$ to a future investigation.


\section{Further Remarks}

\setcounter{equation}{0}

\subsection{Applicability of the quadric ansatz to certain forms of equations}

There is no doubt that the method of quadric ansatz is well applicable to an equation of the `standard' form (\ref{eqforquadric_comp}),
\[\frac{\p}{\p x^i} \left(\, b^{ij}(u) \, \frac{\p u}{\p x^j}  \, \right)  = 0.\] 
We have demonstrated that the method can also be applied to 
 an equation of the form (\ref{modeqforquadric}),
\[\frac{\p}{\p x^i} \left(\, b^{ij}(u) \, \frac{\p u}{\p x^j}  \, \right) + T(\boldsymbol{x}, u, u_{x^i}) = 0,\]
where our generalised procedure of Proposition \ref{prop_modquadric} adds in condition (\ref{modquadredtau}) due to the term $T,$ which puts additional constraints on the matrix ${\bf M}$ of the ansatz.  

Although depending on the term $T,$ we find in our examples of Section $3$ and $4$ that the complicated form of the constraints  makes the general analysis of all solutions constant on a central quadric impractical.  Nevertheless, the method can be used to find particular classes of solutions.

In this section, we suggest a form of the additional term $T$ in (\ref{modeqforquadric}) such that (\ref{modquadredtau})
 does not yield any additional constraint on ${\bf M}.$  Thus the quadric ansatz method is as effective to use with this type of equations as with an equation of the form (\ref{eqforquadric_comp}).   This term $T$ is given by
\be \label{simpleT} T =  m \, b^{ij}(u) \, \frac{\p u}{\p x^i} \, \frac{\p u}{\p x^j},\ee
where $m$ is a constant.  Then, using (\ref{u_xk}), it can be shown that  $\tau  = m g.$  Hence
 (\ref{modquadredtau}) is satisfied by (\ref{modquadredM}), and the quadric reduction is given solely by the pair (\ref{modquadredM}, \ref{modquadredg}) or (\ref{modquadredN}, \ref{modquadredNg}), where 
 system (\ref{modquadredN}, \ref{modquadredNg}) becomes
\be \label{simpmodquredsys}  g \dot {\bf N} = {\bf b}, \qquad g^2  e^{2mu} \det {\bf N} = \zeta. \ee

\vspace{0.5cm}

\noindent {\bf Example $\bs{5.1}$} \,  With $T$ given by (\ref{simpleT}), the $SU(\infty)$-Toda equation and the dKP equation are respectively modified to 
\be \label{simpmodToda} u_{xx} + u_{yy} + (e^u)_{zz} + m \, \Big( u_x^2 + u_y^2 +e^u u_z^2 \Big) = 0,\ee
and 
\[u_{xt} -(uu_x)_x -u_{yy} - m(u u_x^2 - u_x u_t + u_y^2) =0.\]
Assuming the matrix ${\bf M}$ in the quadric ansatz is invertible, all quadric ansatz solutions to the above equations 
can in principle be found using (\ref{simpmodquredsys}).

For example, the solutions to (\ref{simpmodToda}) coming from a diagonal
 matrix ${\bf N}$ (\ref{NToda}) with $\eta = 0$ in (\ref{Todaeta}) are determined by
a second order ODE for $X.$
\, Let $v = e^u,$  the ODE for $X(v)$ is given by
\be \label{simpTodaredODE} X'' = \frac{{X'}^2}{X} + (m-1)\frac{X'}{v} + \frac{v^{2m-1}}{2\zeta} X^2,\ee
where $\dsl X' = \frac{dX}{dv},$  etc.  The family of ODEs (\ref{simpTodaredODE}) includes the Painlev\'e III equation for $m=0,$
which corresponds to the reduction of the $SU(\infty)$-Toda equation given in \cite{T95}.  

It would be interesting to see if an equation of the form (\ref{modeqforquadric}) with $T$ (\ref{simpleT}) has any relevance in geometry or other applications.

\subsection{The hyperplane ansatz} \label{sec:plane}

In addition to the central quadric ansatz (\ref{quadricansatz}),
$M_{ij}(u) x^i x^j = C,$ one can of course consider an ansatz of a general quadric form including terms linear in $x^i,$
or a hypersurface ansatz of higher degree polynomials.

Here, we give a summary of the reduction by the hyperplane ansatz of two forms of equations.  The first one is the form
\be \label{eqforplane}
 b^{ij}(u) \, \frac{\p^2 u}{\p x^i \p x^j}  = 0,
\ee
and the second one is the standard form (\ref{eqforquadric_comp}).

Our interest in the form (\ref{eqforplane}) comes from an observation that two EW equations are of this form.
 The first one is (\ref{dKPEWeq}) with $n=2,$ 
\be \label{dKPEWeq4d} u_{xt} - uu_{xx} - u_{yy} - u_{zz} = 0,\ee
which locally determines an EW structure in four dimensions, where the metric $h$ and the one-form $\nu$ defining the structure is given by
\[h = dy_1^2 + dy_2^2 - 4 dx dt -4 u dt^2, \qquad \nu = -2 u_x dt.\]
The other equation of the form (\ref{eqforplane}) is 
\[u_{xx} + u_{yy} +e^u u_{zz} = 0,\]
which can be shown by direct calculation to be the EW equation for the pair $(h, \nu)$ given by
\[h =e^u(dx^2 + dy^2) + dz^2+ dw^2, \qquad \nu = u_z dz,\]
where $u(x,y,z)$ is independent of one coordinate $w.$

The hyperplane ansatz is given by
\be \label{planeansatz} V_i(u) x^i = C,\ee
where ${\bf V} = (V_i(u))$ is a column vector whose components are functions of $u,$ $C$ is a constant and the summation convention is used as before. The ansatz reduction can be obtained in a similar fashion to the quadric ansatz reduction.  Let us describe this for equation (\ref{eqforplane}).

Assuming the existence of a solution $u(\boldsymbol{x})$ to  (\ref{eqforplane}) given implicitly by (\ref{planeansatz}), implicit differentiation gives
\be \label{u_xkplane} \frac{\p u}{\p x^k} = - \frac{1}{\dot Q} V_k, \ee
where here  $Q(\boldsymbol{x}, u)  := V_{i}(u) x^i$ and 
$\dsl \dot Q := \frac{\p Q}{\p u} =  \dot V_{i} \, x^i.$  \, Then
substituting (\ref{u_xkplane}) into (\ref{eqforplane}) yields
\be \label{matrixeqplane}  \left( 2 ({\bf \dot V}^T {\bf b V}) {\bf \dot V}^T  - ( {\bf V}^T {\bf b V}) {\bf \ddot V}^T  \right) {\bf X} = 0, \ee
where  $\dsl \dot{\bf V}:= \frac{d{\bf V}}{du},$  and ${\bf X}$ is the column vector whose transpose  is the row vector ${{\bf X}^T  = (x^1 \, \dots \, x^N)}$  as before.

Similar to the method of quadric ansatz, one can regard the vanishing of the vector in the bracket of (\ref{matrixeqplane}) as the reduced equation by the hyperplane ansatz.   This gives a system of ODEs for the components $V_i,$ given by 
\be \label{redplane}  \frac{d}{du} (\ln \dot V_i) = 2\frac{{\bf \dot V}^T {\bf b V}}{{\bf V}^T {\bf b V}}, \quad \mbox{for all} \quad i = 1, \dots, N. \ee
Since every component $V_i$ satisfies the same equation (\ref{redplane}), denoting  $V:= V_1,$ it follows that
\[V_i = m_i V + n_i,   \quad \mbox{for all} \quad i = 2, \dots, N,\]
where $m_i$ and $n_i,$ $ i = 2, \dots, N,$ are arbitrary constants.  We now state the ODE for $V$ in the following proposition.
\begin{prop} \label{prop_plane}
Given a PDE of the form {\rm(\ref{eqforplane})},
\[ b^{ij}(u) \, \frac{\p^2 u}{\p x^i \p x^j}  = 0, \quad i, j = 1, \dots, N,  \]
a function $u(\boldsymbol{x})$ defined implicitly by {\rm(\ref{planeansatz})},  $V_i(u) x^i = C,$ is a solution of {\rm(\ref{eqforplane})} if the functions $V_i$ for $i = 2, \dots, N$ are given by
$V_i = m_i V + n_i,$ where $\dsl V:=V_1,$ $m_i$ and $n_i$ are arbitrary constants, and $V(u)$ satisfies the second order ODE 
\[ \ddot V = 2 \, \frac{F V + G}{F V^2 + 2 G V+H} \, \dot V^2,\]
with 
\[F =b^{ij}m_i m_j, \quad G = b^{ij}m_i n_j, \quad H = b^{ij}n_i n_j,\]
$m_1 = 1$ and $n_1=0.$  The solution $u(\boldsymbol{x})$ is thus given by a function of one variable $\dsl w:=\frac{C - n_i x^i}{m_j x^j}.$
\end{prop}

\vspace{1cm}

For an equation of the form (\ref{eqforquadric_comp}), 
\[\frac{\p}{\p x^i} \left(\, b^{ij}(u) \, \frac{\p u}{\p x^j}  \, \right) = 0,\]
the reduction follows similarly, with (\ref{matrixeqplane}) replaced by
\[ \left( \frac{d}{dt}({\bf V}^T {\bf b V}) \, {\bf \dot V}^T  - ( {\bf V}^T {\bf b V}) {\bf \ddot V}^T  \right) {\bf X} = 0.\]
The reduced ODE turns out to be the generalised Riccati equation (\ref{redplaneV2}).
\begin{prop} \label{prop_plane2}
Given a PDE of the form {\rm(\ref{eqforquadric_comp})}, 
\[\frac{\p}{\p x^i} \left(\, b^{ij}(u) \, \frac{\p u}{\p x^j}  \, \right) = 0,\]
a function $u(\boldsymbol{x})$ defined implicitly by {\rm(\ref{planeansatz})},  $V_i(u) x^i = C,$ is a solution of {\rm(\ref{eqforquadric_comp})} if the functions $V_i$ for $i = 2, \dots, N$ are given by
$V_i = m_i V + n_i,$ where $\dsl V:=V_1,$ $m_i$ and $n_i$ are arbitrary constants, and $V(u)$ satisfies the first order ODE 
\be \label{redplaneV2} \dot V = F \, V^2 + 2 G \, V+H,\ee
with 
\[F =b^{ij}m_i m_j, \quad G = b^{ij}m_i n_j, \quad H = b^{ij}n_i n_j,\]
$m_1 = 1$ and $n_1=0.$  
\end{prop}


\section*{Acknowledgements}

The author would like to thank Maciej Dunajski for suggesting the problem of exploring a generalisation of the method of quadric ansatz and for valuable comments and suggestions.  She is also grateful to Paul Tod for helpful correspondence.   





\end{document}